\documentclass{aastex}          
\usepackage{spr-astr-addons}    

\begin{document}
%
\title{Bayesian model-independent evaluation of  expansion rates of the universe}

\shorttitle{Expansion rates of the universe}
\shortauthors{Moncy V. John}

\author{Moncy V. John\altaffilmark{}} 

\altaffiltext{1}{Department of Physics, St. Thomas College, Kozhencherry - 689641, Kerala, India.}

\begin{abstract}
Marginal likelihoods for the cosmic expansion rates are evaluated using the  `Constitution'  data of 397 supernovas, thereby updating the results  in some previous works. Even when beginning with a very strong prior probability that favors an accelerated expansion, we obtain a marginal likelihood for the deceleration parameter $q_0$ peaked around zero in the spatially flat case.  It is also  found that the new data  significantly constrains the cosmographic expansion rates, when compared to the previous analyses. These results may strongly  depend on the Gaussian prior probability distribution chosen for the Hubble parameter represented by $h$, with $h=0.68\pm 0.06$. This and similar priors for other expansion rates were deduced from  previous data. Here again we perform the Bayesian model-independent analysis  in which  the scale factor is expanded into a Taylor series in time about the present epoch. Unlike such Taylor expansions in terms of redshift, this approach has no convergence problem.

\end{abstract}


\section{Introduction}
It is generally accepted that a more appropriate way to measure the acceleration of  expansion of the universe is to resort to a cosmographic or model-independent analysis. In the conventional model-based analyses of distance modulus-redshift ($\mu-z$) data of Type Ia supernova (SN Ia),  the accelerated expansion of the universe is  an indirect inference based on the best fit values of parameters, such as the density parameters $\Omega_m$,  $\Omega_{\Lambda}$, etc.   On the other hand, in a model-independent approach, the scale factor $a(t)$ is  expanded as a Taylor series in time about the present epoch  and the marginal likelihoods of its coefficients are computed using the data. The marginal likelihood for the deceleration parameter gives an estimate of the acceleration of cosmic expansion. Since practically one has to truncate the series to some finite order, the basic assumption here is that $a(t)$ is expressible as a truncated Taylor series or polynomial. Evaluating the deceleration parameter  by adopting this method, it was  confirmed model-independently that the  universe is undergoing an accelerated expansion \citep{mvj1,mvj2}. 

  In this paper we report the   updating of the  marginal likelihood for each of the expansion coefficients found in the above work. This is performed for the case of a fifth order polynomial.  A notable result in the present Bayesian model-independent analysis is that even when beginning with a very strong prior probability that favors an accelerated expansion, the marginal likelihood for the deceleration parameter $q_0$ is found peaked around $q_0=0$ in the spatially flat case.  It is also  found that the new data  significantly constrains the cosmic expansion rates appearing in the Taylor expansion, when compared to the previous data. We also note that successive terms in the series  decreases sufficiently fast, thereby verifying  the assumption of a converging  Taylor series in time for the cosmic scale factor.

Other model-independent approaches, which  Taylor expand  the distance modulus $\mu$ in terms of redshift $z$, have also gained  attention in recent years [See for eg. \citep{turner,visser,lima,seikel}]. But a drawback of this method is that, in principle, it converges only for $\mid z\mid <1$ \citep{visser,lima}. The argument behind this assertion is as follows: For an expanding universe, $z<0$ corresponds to the future and $z=-1$ is the redshift when the universe has expanded to infinite size. Since $z=-1$ is a pole, by standard complex variable theory, the radius of convergence of a series about $z=0$ is atmost $\mid z\mid =1$, so that it fails to converge for $z>1$. When compared to this,  our approach  of expanding the scale factor  in  terms of $t$ about the present epoch $t_0$ is advantageous, for the series converges for all times. Even the lookback time $T\equiv t-t_0$ is evaluated by numerically solving an equation which involves a  Taylor series in time. Hence there is no convergence problem  in the present work. However, it should be noted that all analyses which make use of such Taylor expansions, in practice, employ polynomials and hence convergence is not a serious problem. For instance, one can see that there is  convergence  in certain special cases of the low order polynomial fit by \cite{lima}.

\section{Marginal likelihoods for the cosmic expansion rates}

With $t-t_0\equiv T$, where $t_0$ is the present time, the scale factor of the universe is expanded into a  Taylor series about the present epoch $t_0$ as \citep{mvj1,mvj2}

\begin{eqnarray}
&  a(t_0+T) = a_0 \times   \\ \nonumber
&  \left[1+H_0T-\frac{q_0H_0^{2}}{2!}T^2 
  +\frac{r_0H_0^3}
                    {3!}   T^3  
   -\frac{s_0H_0^4}{4!}T^4 +\frac{u_0H_0^5}{5!}T^5+.. \right] 
         \label{eq:taylor}
\end{eqnarray}
Note that $T$ assumes negative values for the past. For a $k=0$ flat universe, one can put $a_0=1$.  The remaining parameters in the theory are   the Hubble parameter $H_0\equiv 100 h$ km s$^{-1}$ Mpc${-1}$ , the deceleration parameter $q_0$ and higher order expansion rates such as $r_0$, $s_0$, $u_0$, etc. Our task is to deduce the values of these parameters  from the observational data.

For  a light pulse   emitted  from a SN situated at the coordinate $r_1$ at time $t_1$ and reaching us at $r=0$ at time $t_0$,  the RW metric allows one to write

\begin{equation}
\int_{t_1}^{t_0} \frac{cdt}{a(t)} = \int_{r_1}^{0} \frac{dr}{(1-kr^2)^{1/2}} \label{eq:integral}.
\end{equation}
For a $k=0$ RW metric,  this  can be used
 to obtain

\begin{equation}
r_1= \int_{t_1}^{t_0} \frac{cdt}{a(t)}=\int_{T_1}^{0}\frac{dT}{a(t_0+T)}.
\end{equation}
 With this, we can
 compute the luminosity distance $D=r_1a_0(1+z)$.
An important part of the calculation is the solution of the  equation 
\begin{equation}
 1+z=\frac{a(t_0)}{a(t_0+T_1)}, \label{eq:numsolzT}
\end{equation} 
used to find $T_1$ in terms of $z$, for each combination of parameter values. This is done in a direct and purely numerical way.

We may now obtain the distance modulus as $\mu=5 \log \left({D}/{1 \hbox {Mpc}}\right)+25$.
Here $D$ and hence $\mu$ are functions of  $z$ and  contain parameters  
$h$,
$q_0$,
$r_0$, $s_0$, $u_0$, etc. In the following, we keep only terms up to fifth order in the  Taylor expansion.

The likelihood function is  

$$
{\cal L}= \exp[-\chi^2(h,q_0,r_0,s_0,u_0)/2]
$$
where $\chi^2$ is given by

\begin{equation}\label{eq:chi2}
\chi^2 = \Sigma _k \left( \frac{\hat{\mu}_k -\mu_k(z_k;h,q_0,r_0..)}{\sigma_k}\right)^2.
 \end{equation}
 Here $\hat{\mu}_k$ is the measured value of the distance modulus  of the $k^{th}$ supernova, $\mu_k(z_k;h,q_0,..)$ is its expected value (from theory) and $\sigma _k$ is the uncertainty in the measurement.

The likelihood  for the truncated Taylor series form of scale factor can be found  as

\begin{eqnarray} \nonumber
& {\cal L}(M) = \int dh  \int dq_0 \int dr_0 \int ds_0 \int du_0 \\
&p(h) p(q_0)p(r_0)p(s_0)p(u_0)\; 
e^{-\chi^2 /2} . \label{eq:likelihood}
\end{eqnarray}
where $p(h)p(q_0)p(r_0)p(s_0)p(u_0)$  is a product of Gaussian  probability distributions  of each of the parameters. This is an approximation to $p(h,q_0,r_0.....)$, the  prior probability distribution to be introduced  in equation (\ref{eq:likelihood}).

The marginal likelihood of any one parameter can be computed by integrating the likelihood function, multiplied by an appropriate  prior probability distribution, over all  parameters except  the concerned one.  For instance, the marginal likelihood for $q_0$ can be found as

\begin{eqnarray} \nonumber
& {\cal L}(q_0) = \int dh   \int dr_0 \int ds_0 \int du_0 \\
&p(h) p(r_0)p(s_0)p(u_0)\; 
e^{-\chi^2 /2} . \label{eq:marglikelihood}
\end{eqnarray}

A salient feature in the present computation  is that while using (\ref{eq:marglikelihood}), the marginal likelihoods obtained in the previous analysis \citep{mvj1,mvj2} are taken as the prior probability distributions, for the corresponding coefficients.  These references  have used  flat priors, since there were no other previous work evaluating these marginal likelihoods. But  it was proposed there itself that the posterior marginal likelihoods obtained in it shall be used as priors for future analyses. We note that the present work is the appropriate place to make use of this. However, it would not be computationally feasible  to  use  the posterior in the previous analysis  as prior, in terms of a table of values. Therefore,  we approximate those  distributions by Gaussian functions, with the corresponding mean and standard deviations obtained in \citep{mvj2}. A comparison with the actual plots show that this is a reasonable approximation for most coefficients. The product of such individual priors is the combined prior,  used in equation (\ref{eq:marglikelihood}). 

It should  be noted that the ranges of flat priors in the previous analysis were chosen on the basis of the same `All SCP' SN data in \cite{perl1} itself, eventhough it ran the risk of using the same data twice. There we first found the ranges  of  the `contributing' values of the parameters by  varying them arbitrarily. Later,  flat prior probabilities were assigned for those ranges  and they were  used to find the posterior marginal likelihoods. In the present analysis, the  marginal likelihoods thus obtained are used as priors. 

 As mentioned above, the present model-independent analysis uses  the `Constitution'  data \citep{hicken} of 397 SN. In this connection, it shall be noted that  some SN Ia, which appeared in the dataset of 54 `All SCP' SN used in the previous analysis are present in the  Constitution set too. But the values of $\mu_k$, $z_k$ and the errorbars  of such objects are found  modified to some extent in the new release. Therefore, it was  opted not to exclude such SN from the Constitution set. We note that  at any rate,   the Gaussian  priors obtained from the previous analysis  are a better option than flat priors.

 We have computed the marginal likelihoods of four important   expansion rates of the present universe, namely  $q_0$, $r_0$, $s_0$ and $u_0$, and the results are shown in Figs. (1)-(4). Terms up to fifth order are kept in the expansion, but only the flat ($k=0$) case is considered. This is equivalent to assuming a $\delta$-function prior for the flat spatial geometry. 
The joint prior probability used for other parameters was  the product of individual Gaussian functions in each parameter with mean and standard deviations as  follows: 
 $h=0.68\pm 0.06$, $q_0=-0.90\pm 0.65$, $r_0=2.7\pm 6.7$, $s_0=36.5\pm 52.9 $, and $u_0=142.7\pm 320$ \citep{mvj2}. In each case, the integrations were performed in the 2$\sigma$ range of each of the parameters. We have performed variation with respect to $h$, though marginal likelihood for this parameter was not drawn. The step sizes chosen for these parameters were $\Delta  h=0.01 $, $\Delta q_0 =0.1 $, $\Delta r_0 =1 $, $\Delta s_0 =20 $ and $\Delta u_0 = 100$.

\begin{figure} [t]
\plotone{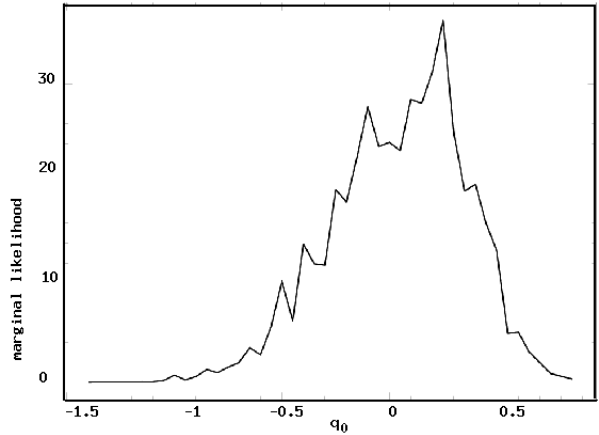}
\caption[]{
Marginal likelihood for the parameter $q_0$ (in units of $10^{-105}$), while using the polynomial of order 5
}
\label{fig:fig1}
\end{figure}

\begin{figure} [t]
\plotone{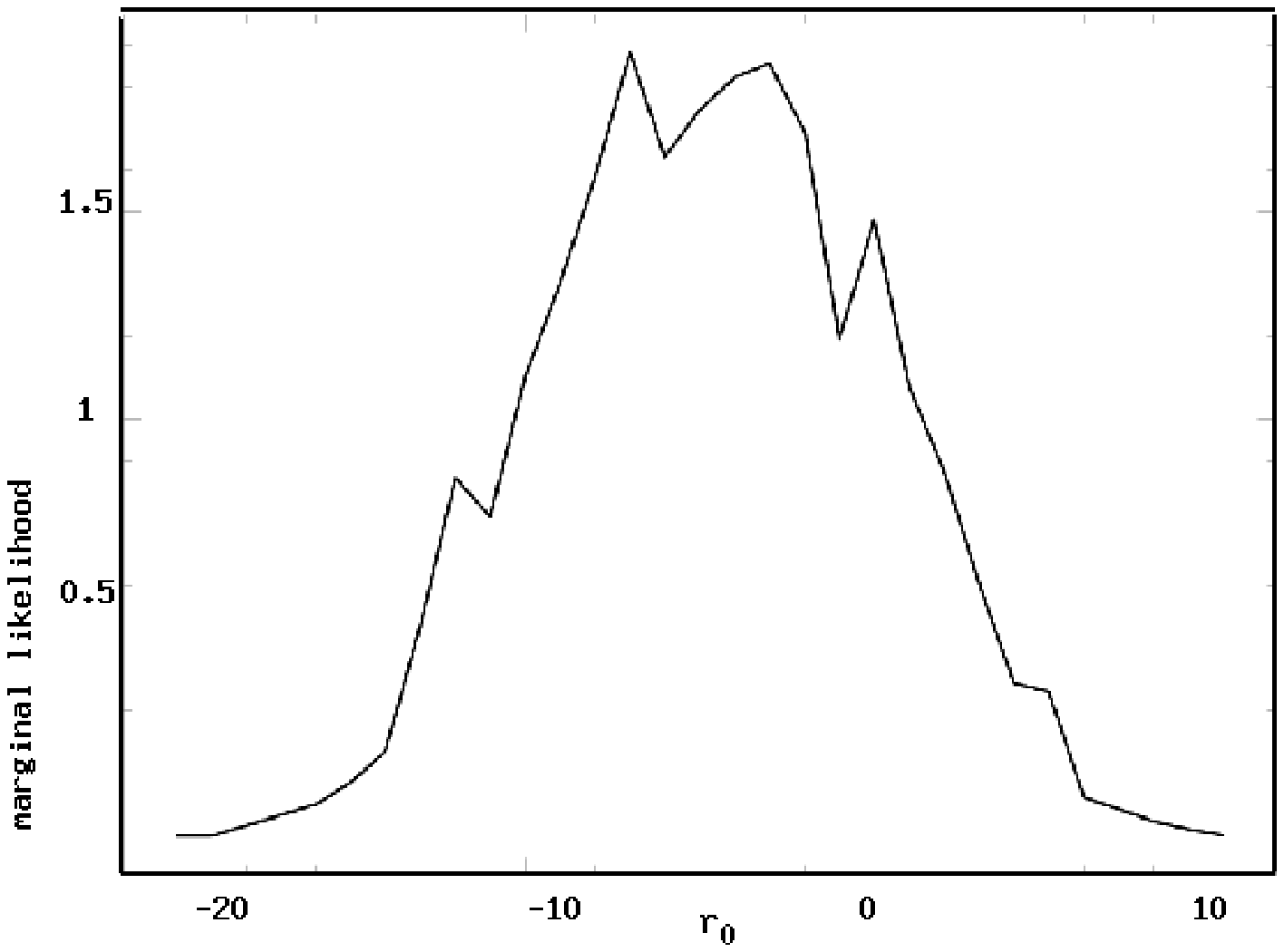}
\caption[]{
Marginal likelihood for the parameter $r_0$ (in units of $10^{-105}$), while using the polynomial of order 5
}
\label{fig:fig2}
\end{figure}

\begin{figure} [t]
\plotone{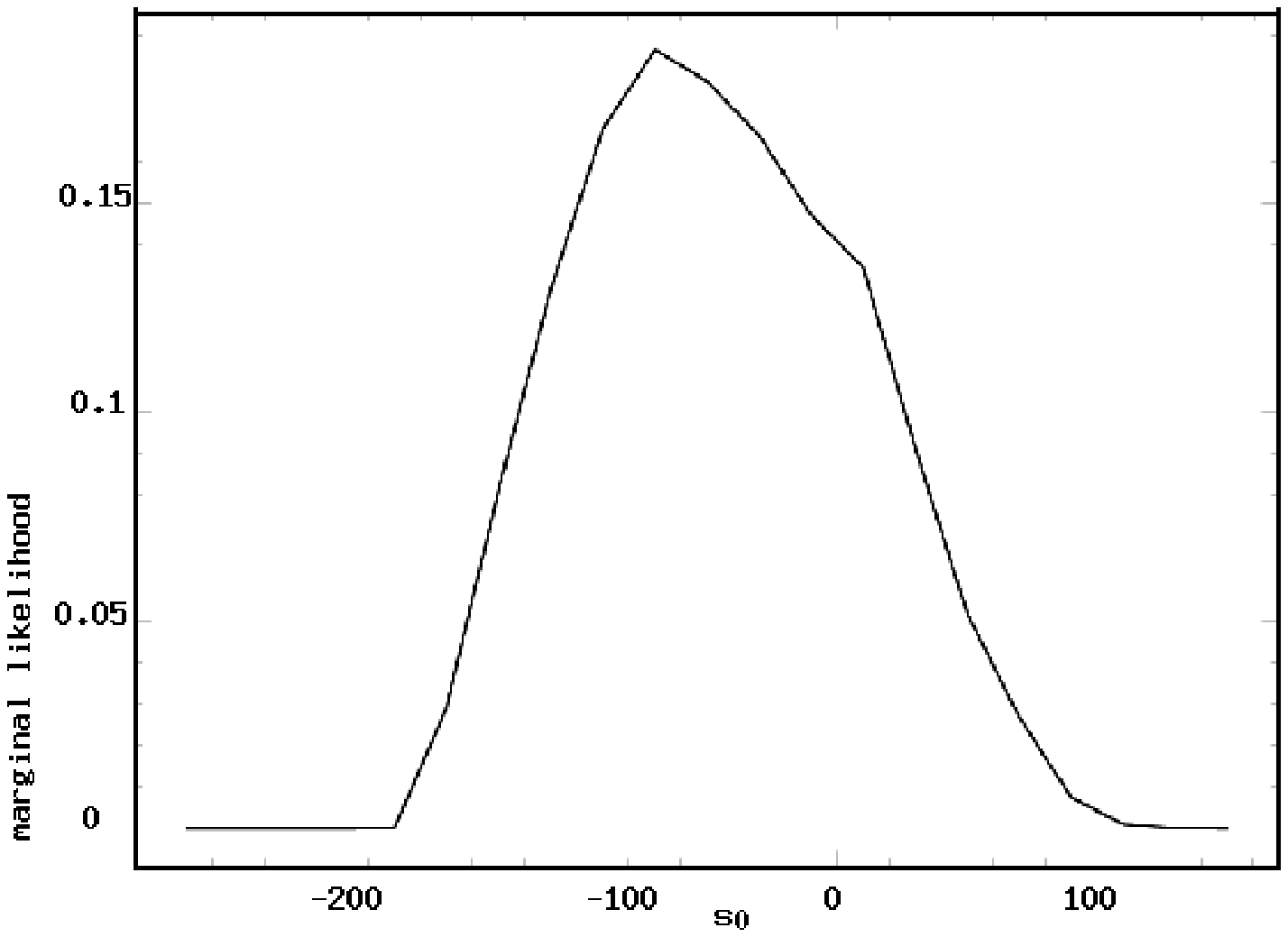}
\caption[]{
Marginal likelihood for the parameter $s_0$ (in units of $10^{-105}$), while using the polynomial of order 5
}
\label{fig:fig3}
\end{figure}

\begin{figure} [t]
\plotone{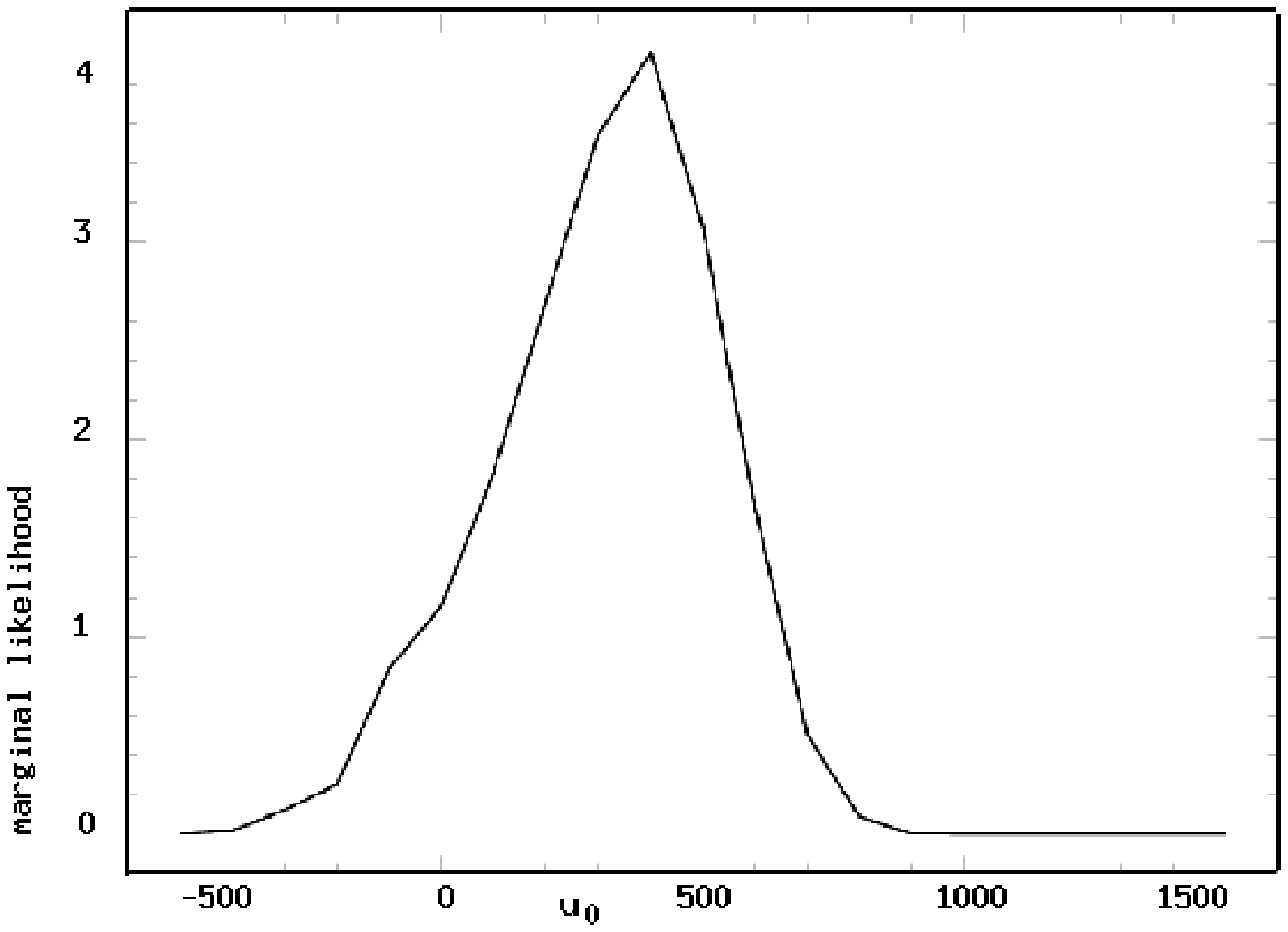}
\caption[]{
Marginal likelihood for the parameter $u_0$ (in units of $10^{-107}$), while using the polynomial of order 5
}
\label{fig:fig4}
\end{figure}

\begin{figure} [t]
\plotone{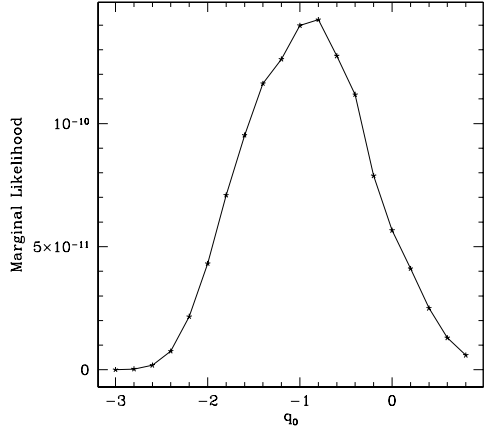}
\caption[]{
Marginal likelihood for the parameter $q_0$, obtained in \cite{mvj2}, while using the polynomial of order 5 and the data in \cite{perl1}.
}
\label{fig:fig5}
\end{figure}

The results  show that there is  significant constraining of the parameters  while using the new and refined data,  compared to the corresponding results in \citep{mvj2}. It is to be reminded that the marginal likelihoods are not precisely probability distributions for the parameters; instead, they are the probability for the data, given the model and the parameter values. However, we here compute mean and standard deviations, considering the marginal likelihoods as distributions. The new mean and standard deviations are the following:  $q_0=0.04\pm 0.30$, $r_0=-4.5\pm 4.6$, $s_0=-42.8\pm 52.5 $, and $u_0=320.5\pm 213.0$. The marginal likelihood for $q_0$ obtained by \cite{mvj2}, which is also used as prior for this parameter in the present work,  is reproduced here in Fig. (\ref{fig:fig5}) for comparison. The standard deviations of each of these parameters, except that of $s_0$, have decreased substantially and this leads to the above assertion that the Constitution data constrains the cosmic expansion rates significantly.

It shall be noted that even when beginning with a  prior probability distribution centred around $q_0=-0.9$, which is strongly in favor of an accelerated expansion, we ended up with a marginal likelihood peaked around $q_0\approx 0$. Thus whereas the data in \citep{perl1}  validated the claim of accelerated expansion, the Constitution SN  dataset in \cite{hicken} favors a coasting evolution; i.e.,  the universe  may  neither be accelerating nor decelerating.  However, the presence of substantial amount of dark energy and dark matter would still be required to explain the data.

Here one observes that the result $q_0\approx 0$ could be connected to the value of $h$ and also that properly including $h$ in the analysis may further decrease the constraining power of the data. Therefore one should explore the consequences of using Gaussian priors on $h$  from other measurements too. In fact,  we have marginalised the likelihoods over the Hubble parameter, with a Gaussian prior $h=0.68\pm 0.06$, as mentioned above. But since the likelihood curve for $h$ obtained from the previous analysis by \cite{mvj2} is not very sensitive to its variation  in the concerned range (unlike the case of $q_0$, $r_0$, etc.), it is more appropriate to employ priors for $h$ deduced from other measurements. We propose that this procedure shall be followed in future analyses.

The considerable spread left in the marginal likelihoods  shows that even now there is  some freedom in choosing the values of those parameters.  In other words,  there is a sizable volume in the parameter space that can have the same low $\chi^2$. But  this should not be viewed as a drawback of the analysis; instead, this simply reflects the fact that the data are  not yet accurate enough.
Some recent analyses of Constitution SN data endorses this result  \citep{varun}.   This freedom in SN data was noted  earlier  \citep{mvj1,mvj2}, which highlights the strength of the Bayesian model-independent approach.

Based on the mean values obtained for these parameters, we compute the successive terms in the series (\ref{eq:taylor}). With time in units of $10^{17}$ s, the series can be written as

\begin{eqnarray} \nonumber
 1+2.106\times 10^{-1} T-2.22\times 10^{-2} q_0 T^2 \\ \nonumber
+1.55\times 10^{-3}r_0 T^3 - 0.819\times 10^{-4} s_0 T^4 \\ \nonumber
+3.45\times 10^{-6} u_0 T^5+......,
  \label{eq:converge}
\end{eqnarray}
where we have taken $h=0.65$ (only to evaluate this series). With the values of the parameter in the ranges obtained in the analysis, this series appears to converge even for  $\mid T \mid$ as large as $\approx 3\times 10^{17}$ s. However, this feature is not  essential for our analysis, for we have assumed only a polynomial form for the scale factor. The situation was not different in the previous work either.

\section{Conclusion}

We assumed that  a Taylor series form for the scale factor $a(t)$  is valid and  attempted to find the coefficients in this expansion using the recent Constitution SN data. The new marginal likelihoods obtained for its coefficients  give valuable information regarding the expansion history of the universe. It is found that there is significant constraining of these parameters when compared to  previous analyses using the  data in \citep{perl1}. 
The shift in the computed mean value of the deceleration parameter $q_0$, from that found in the previous analysis is noteworthy. Even when we start with a  prior probability distribution that strongly favors an accelerating universe,   the marginal likelihood for the deceleration parameter obtained from the  present analysis using the Constitution data is found peaked around $q_0=0$.   However, we reiterate that the considerable spread still found in the likelihoods of these parameters indicate  freedom in the choice of their numerical values.

A distinguishing feature of our analysis is that the marginal likelihoods for each parameter obtained in the previous case is chosen as the prior probability distribution in the present one, thereby implementing the Bayesian method in true spirits. The work is also intended as a demonstration of this fundamental requirement in Bayesian analysis. However, we have noted that the results obtained in this paper may heavily depend on the prior chosen for $h$. Thus it is important to evaluate expansion rates using prior for $h$ deduced from other measurements too. It is expected that in  future when the SN dataset becomes large enough, the expansion coefficients  get sharply peaked marginal likelihoods and  become the most basic model-independent description of the expansion history of the universe.

\acknowledgments{
It is a pleasure to thank Professor J. V. Narlikar for helpful discussions. The author also wishes to thank IUCAA, where most of these computations were done, for hospitality during a visit under the associateship program and the University Grants Commission (UGC) for a research grant under MRP.}


\begin{thebibliography}{}
\bibitem[Cattoen \& Visser (2007)]{visser}  Cattoen, C., \&  Visser, M., 2007, preprint, gr-qc/0703122v3
\bibitem[Guimaraes et al. (2009)]{lima}  Guimaraes, A. C. C.,  Cunha, J. V., \&  Lima, J. A. S. 2009  JCAP  10, 010
\bibitem[Hicken et. al. (2009)]{hicken}   Hicken, M., et. al.,  2009 \apj 700, 1097 
\bibitem [John (2004)]{mvj1} John, M. V. 2004, \apj, 614, 1
\bibitem [John (2005)]{mvj2} John, M. V. 2005, \apj, 630, 667
\bibitem[Knop et. al. (2003)]{perl1} Knop, R. A., et. al., 2003, \apj, 598, 102
\bibitem[Seikel \& Schwarz (2009)]{seikel}  Seikel, M., \&  Schwarz,  D. J. 2009   JCAP 2, 024 
\bibitem[Shafieloo et.al. (2009)]{varun}   Shafieloo, A.,   Sahni, V., \&   Starobinski, A. A.    2009 Phys. Rev. D 80, 101301
\bibitem[Shapiro \& Turner (2006)]{turner}  Shapiro, C., \&   Turner, M. S. 2006 \apj     649, 563
\end{thebibliography}
\end{document}